\documentclass[%
 reprint,
 amsmath,amssymb,
 aps
]{revtex4-1}

\usepackage{braket}
\usepackage{graphicx}
\usepackage{dcolumn}
\usepackage{bm}
\usepackage{xcolor}

\begin{document}
--

\title{Imprinting chirality on atoms using synthetic chiral light fields}

\author{Nicola Mayer$^{1}$,
Serguei Patchkovskii$^{1}$,
Felipe Morales$^{1}$,
Misha Ivanov$^{1,2,3}$,
Olga Smirnova$^{1,4}$}

\affiliation{
$^1$Max-Born-Institute, Max-Born str. 2A, 12489 Berlin, Germany \\
$^2$Department of Physics, Humboldt University, Newtonstr. 15, D-12489 Berlin, Germany\\
$^3$Blackett Laboratory, Imperial College London, SW7 2AZ London, United Kingdom.\\
$^4$ Department of Physics, Technical University Berlin, Berlin, Germany}

\date{\today}

\begin{abstract}
Atoms are usually thought of as achiral objects. 
However, one can construct superpositions of atomic states that are chiral
\cite{Ordonez:2019aa}. 
Here we show how to excite such superpositions with tailored light fields 
both in the weak-field and strong-field regimes, using realistic laser parameters. First, we  
use time-dependent Schrodinger equation (TDSE) simulations to demonstrate the creation of a time-dependent bound chiral wavepacket in sodium atoms. 
Second, we show how the time-dependent handedness 
of this wavepacket can be probed by photoelectron circular dichroism, in spite of the central symmetry
of the core potential.  Third, we use TDSE simulations to show how chirality can be  directly imprinted on a photoelectron wavepacket created 
by strong-field ionization and introduce an unambigous chiral measure that allows us to characterize its handedness.
\end{abstract}

\maketitle

An object is chiral when it cannot be superposed on its mirror reflection. This geometric property is found at all scales, 
from galaxies to elementary particles, and plays a key role at atomic and molecular scales. 
The importance of molecular chirality in chemistry and biology has led to the development of a plethora of methods for 
studying chiral molecules and distinguishing the enantiomers -- the mirror-image molecular twins. 
Among these methods, the photoelectron circular dichroism (PECD), from single-photon \cite{Bowering:2001aa} to multi-photon 
\cite{Lux:2012aa} to strong-field \cite{Beaulieu:2016aa} regime, stands out due to its very high sensitivity to
molecular chirality. Its recent extension to the attosecond ($10^{-18}$ s) regime \cite{Beaulieu:2017aa} is allowing one to 
study chiral behaviour at the electronic time-scale. 

The PECD owes its high sensitivity to the fact that it works in the electric dipole approximation, in 
contrast to such popular methods as absorption circular dichroism (CD) or optical rotation \cite{Condon:1937aa} 
which require weak magnetic interactions.  The PECD
uses a circularly polarized (CP) field to generate a photoelectron current orthogonal to the polarization plane. 
The scalar product between the helicity of the pulse $\mathbf{E}^*\times\mathbf{E}$ and the direction 
defined by the detector axis $\hat{z}$ yields a pseudoscalar $(\mathbf{E}^*\times\mathbf{E})\cdot\hat{z}$ 
characterizing the handedness of the setup \cite{Ordonez:2018aa}, against which the handedness of the 
molecule is checked. 

It is possible to push enantio-sensitive techniques even further \cite{Ayuso:2019aa,Khokhlova:2021aa,Ayuso:2021aa,Neufeld:2019aa,Cireasa:2015aa,Baykusheva:2018aa}. One way is to use an electric field that is chiral already in the dipole 
approximation \cite{Ayuso:2019aa}, eschewing the need for the chiral measurement setup. This "synthetic chiral light" 
has been shown to lead 
to highly enantio-sensitive High-Harmonic Generation (HHG) \cite{Ayuso:2019aa}, steering of 
Free Induction Decay \cite{Khokhlova:2021aa}, and uni-directional enantio-sensitive light bending  
\cite{Ayuso:2021aa}.

Here, we demonstrate that synthetic chiral light can also be used to imprint chirality on atomic targets. 
While atoms are prototypical achiral objects, we expect the methods described here  to also apply to more complex objects such as achiral molecules. In particular, we show that synthetic chiral light can imprint 
chirality both in the weak-field regime, where we demonstrate the creation of a time-dependent chiral wavepacket in 
Na, and in the strong-field regime, where we demonstrate the creation of chiral photoelectron wavepackets 
and excitation of chiral Freeman resonances.

\begin{figure}
\begin{center}
\includegraphics[width=9cm, keepaspectratio=true]{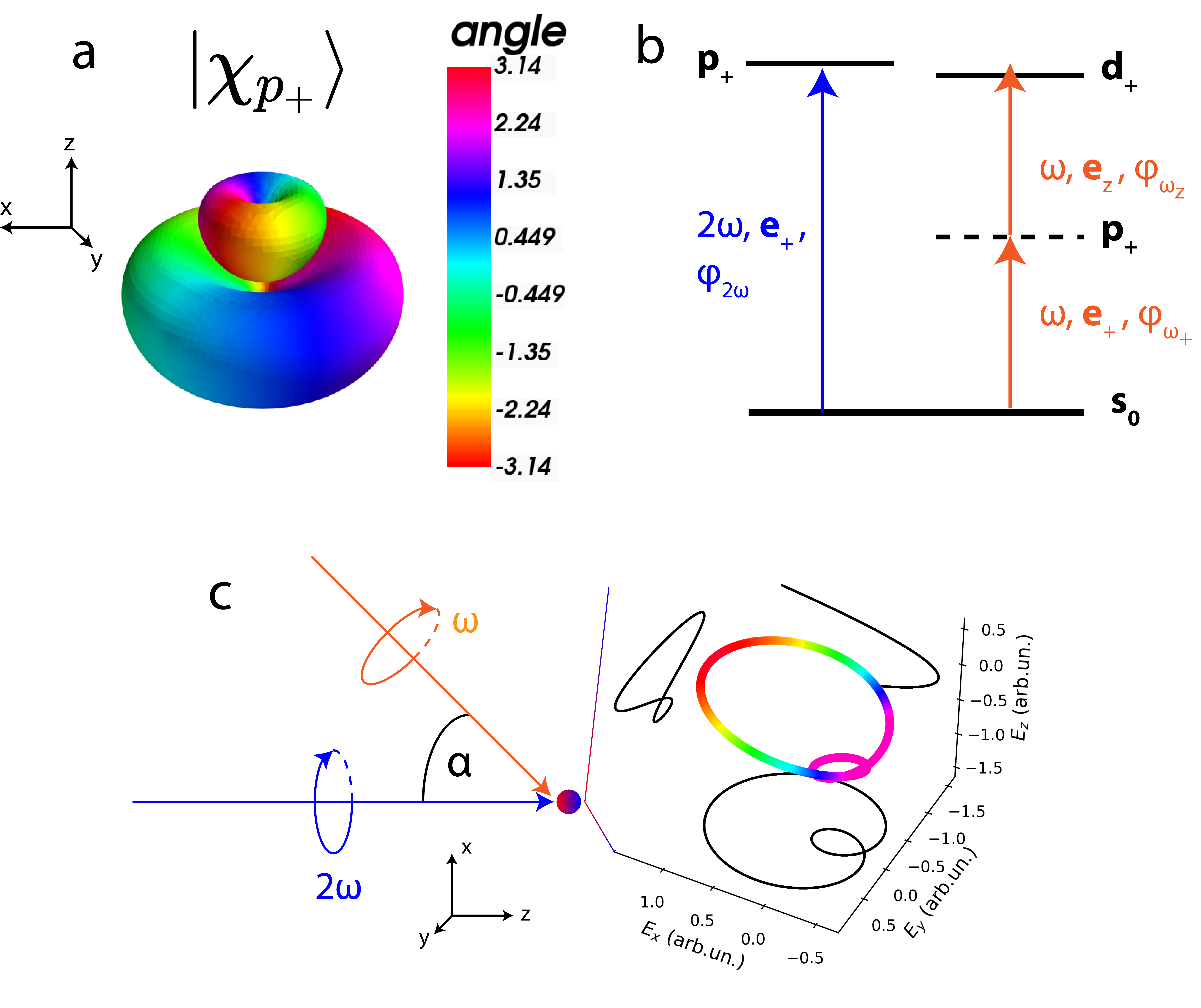}
\caption{\textbf{a)}: isosurface plot of the chiral superposition $|\chi_{p_+}\rangle$, colored according to the phase. \textbf{b)}: photon pathways leading to the excitation of a chiral superposition $|\chi_{p_{\pm}}\rangle$. \textbf{c)}: generation of synthetic chiral light by crossing two co-rotating elliptically polarized beams at $\omega$ and $2\omega$ frequencies at an angle $\alpha$. The figure shows 
the Lissajous curve of the resulting chiral light for $\phi_{\omega,2\omega}=0$, colored according 
to the value of the $E_z$ component, together with its two-dimensional projections.}
\label{Fig1}
\end{center}
\end{figure}

Let us begin with the weak-field case. How can chirality manifest in atoms? An example is the family of chiral wavefunctions presented in  Ref. \cite{Ordonez:2019aa}. Consider one of them -- 
the stationary chiral atomic state
\begin{equation}
|\chi_{p_+}\rangle
=2^{-1/2}
\left(|np_+\rangle+|nd_+\rangle\right)
\end{equation}
where $|np_+\rangle$, $|nd_+\rangle$ are the hydrogenic wavefunctions with principal quantum number $n$, magnetic 
quantum number $m=1$ and angular momentum $\ell=1$ and $\ell=2$ respectively. The isosurface 
$|\chi_{3p_+}|=10^{-6}$, colored according to the phase $\arg[{\chi_{p_+}}]$, is shown in Figure \ref{Fig1}a). 
The superposition of two states with opposite parity breaks the inversion symmetry, leading to the polarization of the 
wavefunction along the quantization $z$ axis, while $m=1$ implies a probability current orbiting in the 
clockwise sense in the xy plane. The combination of these two features results in a chiral wavefunction, whose 
corresponding enantiomer $|\chi_{p_-}\rangle$ can be found by changing the relative phase between the 
two states by $\pi$, i.e. by inverting the direction of polarization of the wavefunction along $z$. In spite of the 
central symmetry of the core potential, one-photon ionization of this oriented chiral state by a field circularly polarized (CP) in the xy plane 
displays a forward/backward asymmetry of photoelectron emission \cite{Ordonez:2019aa}: photoelectrons emitted by opposite 
enantiomers will posses opposite momenta along the z axis.

The degeneracy of the eigenstates composing the $|\chi_{p_+}\rangle$ chiral state is not strictly necessary: 
time-dependent superpositions can also be chiral. As the relative phase between the two states evolves on the time-scale 
defined by the energy difference $\Delta$ between the states, the handedness of the time-dependent chiral state 
$|\chi_{p_+}(t)\rangle
=2^{-1/2}
\left(|p_+\rangle+\exp(-i\Delta t)|d_+\rangle \right)$ 
swaps every half-cycle $\pi/\Delta$. Consequently, the photoelectron current generated via
one-photon ionization of this state by a CP field will also swap its direction every half-cycle.

How can such superpositions be excited? 
Consider the photon pathways leading to the excitation of these states, starting 
from the ground state with $\ell=0$. To reach a $|p_+\rangle$ state, we need a right circularly polarized photon 
(RCP) in the xy plane, i.e. $\mathbf{e}_+=(\mathbf{x}-\text{i}\mathbf{y})/\sqrt{2}$. Excitation of a 
$|d_+\rangle$ state requires two photons: one RCP in the xy plane and one linearly polarized along 
the z axis. Choosing the frequency of the RCP photon leading to the $|p_+\rangle$ state to be twice the frequency of the 
two other photons leading to the $|d_+\rangle$ state, we can excite a chiral state $|\chi_{p_\pm}\rangle$, as shown in Figure \ref{Fig1}b). 
The relative phase between the two pathways $\Delta\phi=\phi_{2\omega}-\phi_{\omega,+}-\phi_{\omega,z}$ 
determines the handedness of the chiral superposition (as clear from the standard perturbation theory.)

Thus, we need a light field that includes a RCP component at the frequency $2\omega$, a RCP component at the frequency $\omega$, 
and a z-polarized component at $\omega$. These are exactly the minimal requirements for the 
synthetic chiral light \cite{Ayuso:2019aa}. One way to create such a field is to use non-collinear, elliptically polarized beams carried 
at $\omega$ and $2\omega$ (Figure \ref{Fig1}c). Choosing the laboratory axis z to be along the direction of 
propagation of the $2\omega$ field, we denote by $\alpha$ the crossing angle between the two beams. 
Due to the non-collinearity, the $\omega$ field acquires a forward component proportional to $\sin\alpha$ in this frame. 
The resulting electric field at the point where the two beams cross is
\begin{eqnarray}E_x&=&E_{\omega}\cos(\omega t+\phi_\omega)\cos\alpha+E_{2\omega}\cos(2\omega t+\phi_{2\omega})\\
E_y&=&\epsilon_\omega E_{\omega}\sin(\omega t+\phi_{\omega})+\epsilon_{2\omega} E_{2\omega}\sin(2\omega t+\phi_{2\omega})\\
E_z&=&E_\omega\cos(\omega t+\phi_{\omega})\sin\alpha\end{eqnarray}
where $\epsilon_{r\omega}$ is the ellipticity of the $r\omega$ field. The Lissajous curve drawn by the polarization vector 
over a laser cycle $T=2\pi/\omega$ (Fig. 1c) is chiral. Its handedness is controlled by the relative phase between the two 
colors $\phi_{\omega,2\omega}=2\phi_\omega-\phi_{2\omega}$, and can be maintained globally across the focus as required, 
preserving the field handedness  \cite{Ayuso:2019aa, Khokhlova:2021aa}.
Shifting the relative phase $\Delta\phi$ by $\pi$ changes the handedness of both the field and the chiral wavefunction 
(when the z-polarized and RCP components at $\omega$ have the same phase.) 
Thus, a field chiral in the dipole approximation should imprint chirality on an atom.

We now use this analysis to create a time-dependent chiral wavepacket in a sodium atom. 
We perform the TDSE simulations using the code of Ref. \cite{Patchkovskii:2016aa}, with the 
model potential for sodium from Ref. \cite{Garvey:1975aa}. We use a radial box size of 3190 a.u., with a log-uniform 
grid of 8000 points. The first 10 points on a uniform grid between $0.036$ a.u. and $0.36$ a.u. are followed by 25 points on a 
logarithmic grid between $0.4$ a.u. and $3.93$ a.u., followed by 7965 points on a uniform grid until the box boundary. 
In order to avoid unphysical reflections, we place the complex absorber from \cite{Manolopoulos:2002} at $3157$ a.u. with $32.7$ a.u. width. We use a time step of 0.025 a.u. and include angular momenta up to $\ell_{max}=13$. We tune the pulse parameters to excite a chiral state, composed of the 
$|4p_+\rangle$ and $|3d_+\rangle$ states, via one- and two-photon transitions. The energy difference between the two states 
is such that the chiral wavepacket should evolve on a 38.5 fs period. The chiral pulse has components carried at 
330 nm and 660 nm, with a Gaussian envelope of 15 fs FWHM. The angle between the two beams is $\alpha=10^{\circ}$. 
The $2\omega$ field is RCP in the laboratory xy plane and propagates along the laboratory z-axis; the $\omega$ field is elliptically 
polarized in its local frame with an 
ellipticity $\epsilon_\omega=\cos\alpha$, such that its projection onto the laboratory xy plane yields 
RCP light.
\begin{figure*}
\begin{center}
\includegraphics[width=15cm, keepaspectratio=true]{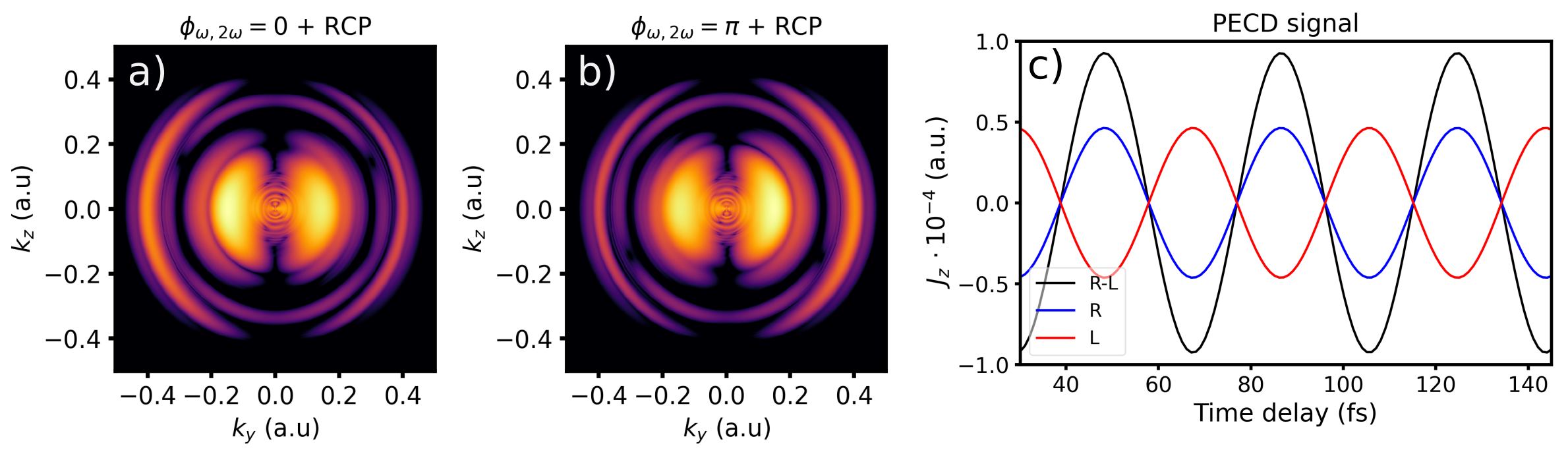}
\caption{Imprinting chirality on a sodium atom in the weak-field regime. Panels (a), (b) show the photoelectron angular distributions for the right ($\phi_{\omega,2\omega}=0$) and left ($\phi_{\omega,2\omega}=\pi$) field enantiomers of the pump pulse respectively. The probe is a RCP pulse carried at $\lambda=$ 400 nm, with pump-probe delay of $30$ fs. Panel (c) shows the pump-probe delay-resolved photoionization current along the z axis in sodium for the right  $I_R$ (blue line) and left $I_L$ (red line) enantiomers, while the black line shows the PECD signal, defined as $I_R-I_L$.}
\label{Fig2}
\end{center}
\end{figure*}
We define the right enantiomer of the field to correspond to $\phi_{\omega,2\omega}=0$, 
and find the left enantiomer by changing $\phi_{\omega}$ by $\pi/2$ ($\phi_{\omega,2\omega}=\pi$). Both fields have equal 
strengths $E=0.005$ a.u. (intensity $I=0.87$ TW/cm$^2$.) 
As a probe, we use a CP 400 nm pulse with strength $E=0.001$ a.u. and a sin$^2$ envelope with 30 fs duration, 
co-propagating with the $2\omega$ field along the z axis. The pump-probe delay is defined with respect to the center of their intensity envelope. In order to retrieve the photoelectron spectrum we project the wavefunction at the end of the pulses on the scattering states using the iSURFC method described in Ref. \cite{Morales:2016aa}. Figs. 2a) and 2b) show the angular- and energy-resolved photoelectron distributions after a RCP probe pulse delayed by $\tau=30$ fs with respect to the chiral pump. In Fig. 2a the relative phase between the two colors forming the chiral pump is $\phi_{\omega,2\omega}=0$, while in Fig. 2b the relative phase between the two colors is $\phi_{\omega,2\omega}=\pi$. One-photon ionization by the probe pulse results in the feature at $k\simeq0.35$ a.u., which displays forward/backward asymmetry with respect to the propagation axis of the probe pulse. In order to better isolate the enantio-sensitive signal, we focus on the photoelectrons with momentum along the z axis and calculate the photoelectron current as
\begin{equation}\label{eq:Jz}J_z(\tau)=\int dk_z\,k_z |a(k_z,\tau)|^2\end{equation}
where $a(k_z,\tau)$ is the amplitude of the photoelectron with momentum $\mathbf{k}=k_z\hat{z}$ and $\tau$ is the relative time-delay between pump and probe. 
Figure \ref{Fig2}c) shows the delay-resolved PECD signal integrated around the feature at $k=0.35$ a.u. resulting from one-photon ionization of the chiral superposition by the 400 nm probe. The PECD signal is defined as $I_{PECD}=I_R-I_L$ (black line in Fig. 2c), where $I_{R/L}$ are the difference signals for the right and left enantiomers respectively (blue and red lines in Fig. 2c) respectively) and $I_{R/L}=I^+_{R/L}-I^{-}_{R/L}$ is the difference signal for ionization by a RCP (+) or LCP (-) probe. The slow modulation 
at the frequency difference between the $3p_+$ and $4d_+$ states of sodium is clearly visible. Our results confirm that synthetic chiral field can imprint chiral quantum states on an atom.
\begin{figure*}
\begin{center}
\includegraphics[width=18cm, keepaspectratio=true]{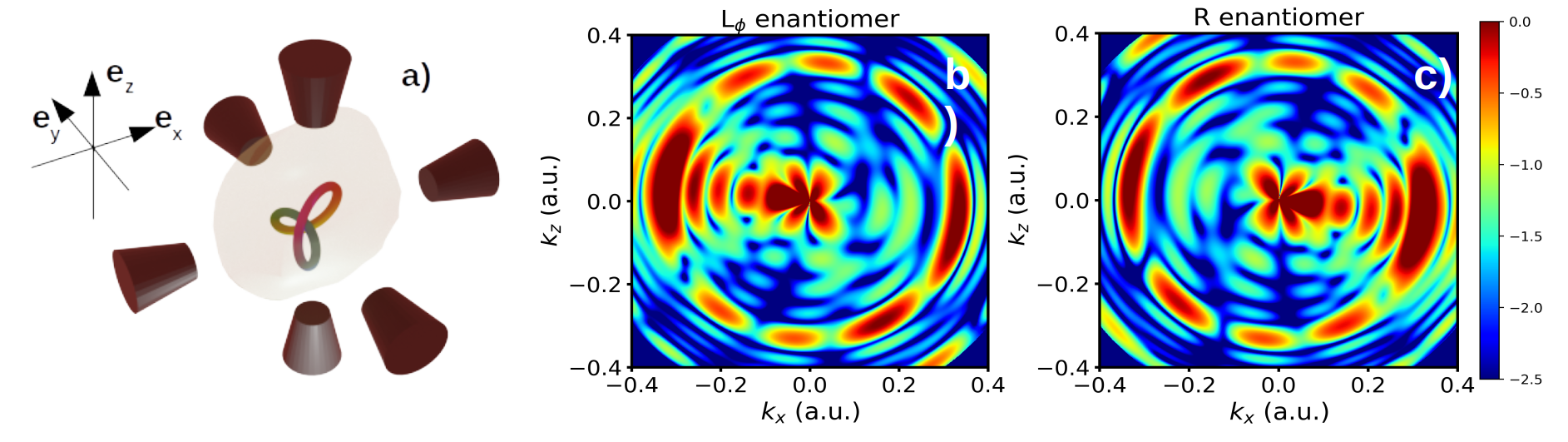}
\caption{Imprinting chirality on an hydrogen atom in the strong-field regime. (a) Chiral setup formed by three sets of detectors placed along orthogonal directions, with handedness characterized by the pseudoscalar $h=\mathbf{e}_x\cdot\left(\mathbf{e}_y\times\mathbf{e}_z\right)$. The Lissajous curve shows the chiral bicircular field, colored according to time. (b,c) Slices through the 3D photoelectron spectrum at $k_y=0$ obtained by TDSE simulations in hydrogen for the left and right enantiomers of the field, on logarithmic scale. The left enantiomer is found by shifting the phase in each beam by $\pi$, corresponding to an inversion. For long pulses, the same result is achieved by changing the relative phase between the two colors by $\pi$.}
\label{Fig3}
\end{center}
\end{figure*}

Let us now consider the strong-field case. 
In the simple model of strong-field ionization,
the final electron momentum $\mathbf{p}$ at the detector is $\mathbf{p}=\mathbf{v}+\mathbf{A}(t_i)$, where $t_i$ is the 
ionization time,  $\mathbf{A}(t)$ is the field vector potential, and $\mathbf{v}\simeq 0$ is the initial velocity of the photoelectron. 
This mapping between the chiral vector potential $\mathbf{A}(t_i)$ and the final electron momentum $\mathbf{p}$ immediately suggests that it should be possible to imprint chirality directly on the photoelectron wavepacket. To characterize such a chiral wavepacket, we can again use a chiral setup \cite{Ordonez:2018aa}, this time offered by the Reaction Microscope (COLTRIMS, see e.g.\cite{DORNER200095}).
Its handedness is defined by the pseudoscalar $h=\mathbf{e}_x\cdot\left(\mathbf{e}_y\times\mathbf{e}_z\right)\neq0$, formed by the three axes $\mathbf{e}_i$ of the 
laboratory frame (Fig. 3a),  in which one measures the photoelectron distribution. 
We can use this setup to obtain and analyze 
the full 3D photo-electron distribution $W(\mathbf{p})=|a(\mathbf{p})|^2$, where $a(\mathbf{p})$ is the amplitude for the photoelectron momentum $\mathbf{p}$ measured
relative to the laboratory frame in Fig. 3(a). 

Given the photoelectron distributions of two enantiomers $W_R(\mathbf{p})$ and $W_L(\mathbf{p})$, we 
can characterize their chirality using the overlap measure \cite{Neufeld:2020aa, DOC:1991aa, Gilat:1989aa, Petitjean:2003aa}
\begin{equation}\chi=\frac{\min_{\rho}\int d\mathbf{p}|W^{\rho}_L(\mathbf{p})-W_R(\mathbf{p})|}{\int d\mathbf{p}W_R(\mathbf{p})}.\end{equation}
where $\rho=(\alpha,\beta,\gamma)$ are the Euler's angles, $W^{\rho}_L(\mathbf{p}_\rho)=W_L(\mathbf{p})$, where $\mathbf{p}_{\rho}=U_{\rho}\mathbf{p}$ and $U_{\rho}$ is the Euler rotation matrix. If a photoelectron distribution is achiral,  its mirror reflection can be superimposed on it. Hence,  $\chi=0$ characterizes an achiral distribution, while $\chi\neq0$ a chiral one. Further details on the implementation of $\chi$ are given in the supplementary material \cite{SuppNotes}.

We verify this proposal  by performing 
TDSE simulations in a hydrogen atom for  
a chiral pulse composed of 400 nm + 800 nm counter-rotating fields, with one-cycle ramp-on and ramp-off and a 4 cycle flat top
(see Fig.3a for the Lissajous curve). Both colors have equal intensity of 31 TW/cm$^2$, the crossing angle 
between the two beams is $\alpha=10^{\circ}$ and we choose the z axis along the propagation 
direction of the $2\omega$ field. The $\omega$ field is elliptically polarized in its local frame so that it appears circular in the xy plane of the laboratory. To change the field enantiomer, we either flip the helicities of the driving pulses or change the relative phase between the two colors $\phi_{\omega,2\omega}$ by shifting the two phases $\phi_\omega$ and $\phi_{2\omega}$ by $\pi$. Note that flipping the field helicities corresponds to reflection along $y$ (or $x$) axis, while the change of phases corresponds to an inversion $\mathbf{r}\rightarrow-\mathbf{r}$. The two ``left-handed'' Lissajous curves are related by rotation around the axis of reflection by $180^{\circ}$. We label the corresponding photoelectron distributions respectively as $W_{L\epsilon}(\mathbf{p})$ and $W_{L\phi}(\mathbf{p})$,
their mirror reflections are labelled as $W^{-1}_{L\epsilon}(\mathbf{p})$ and $W^{-1}_{L\phi}(\mathbf{p})$ 
while $W_R(\mathbf{p})$ corresponds to the right-handed field enantiomer.


We use a radial box size of 1190 a.u., with a log-uniform grid consisting of 3000 points. 
The first 10 points are on a uniform grid between $0.036$ a.u. and $0.36$ a.u., followed by 25 points on a logarithmic grid between $0.4$ a.u. and $3.93$ a.u. and finishing with 2965 points on a uniform grid extending until the box boundary. The complex absorber from \cite{Manolopoulos:2002} is placed at $1158$ a.u. with 32 a.u. width. We use a timestep of 0.0025 a.u. and include angular momenta up to $\ell_{max}=60$. The photoelectron angular distributions are extracted using the iSURFC method \cite{Morales:2016aa}.


The simulations 
yield $\chi(W_R,W_{L\epsilon})=\chi(W_R,W_{L\phi})\simeq0.212$ and $\chi(W_R,W_{L\epsilon}^{-1})=\chi(W_R,W_{L\phi}^{-1})\simeq10^{-8}$. We have also considered numerically inverted photoelectron distributions $W^{-1}_R(\mathbf{p})=W_R(-\mathbf{p})$ and obtained $\chi(W_R,W_R^{-1})=0.212$. These results unambiguously prove the chiral nature of the photoelectron distributions.   
The chiral measure is on the order of the relative strength between the z- and x-y components of the fields. We have also verified that for the achiral pulses contained in the xy plane the photoelectron distributions yield $\chi=0$. Further simulation details can be found in the supplementary material \cite{SuppNotes}.
Our analysis of the chirality of photo-electron distributions
is also applicable to the recent work on a closely related subject \cite{Katsoulis:2021aa}, where globally chiral light  \cite{Ayuso:2019aa} was used to study 3D photo-electron distributions in the strong-field regime.

In Figs. 3(b,c) we show slices through the 3D PES obtained by TDSE simulations at $p_y=0$ for the left (b) and right (c) enantiomers of the field. Their mirror symmetry is clear. Interestingly, a five-fold low-energy structure appears at $p=0.1$ a.u. Its appearance results from multiphoton resonances with Stark-shifted Rydberg states, similar to Ref. \cite{Stammer:2020aa}. 
This suggests that, in the strong-field regime, one should also be able to excite chiral Freeman resonances \cite{Freeman:1987aa}. Thus, while intermediate resonances are not necessary in order to imprint chirality on the photoelectron wavepacket, it is still possible to excite bound chiral wavepackets in the strong-field regime.



Our work opens several avenues for further research. 
For example, instead of using non-collinear setup, 
one can generate synthetic chiral light 
by using collinear but 
tightly focused beams. 
For example, consider two circularly polarized, counter-rotating tightly-focused collinear Gaussian beams carrying orbital angular momentum (OAM) $\ell$ and propagating along the $z_L$ axis of the laboratory;  beams with nonzero OAM
are excellent tools for generating and controlling strong longitudinal fields. 
Using  RCP light carried at 
$\lambda_\omega=660$ nm with OAM $\ell_\omega=1$, and LCP $\lambda_{2\omega}=330$ nm 
with OAM $\ell_{2\omega}=-1$, and assuming focal waist radius
of $W_0=1.5\,\mu$m, the fields acquire a forward component that reaches $\simeq10$\% of the in-plane $\omega$ field ($\simeq5$\% for the $2\omega$ component) along a ring with radius $\rho\simeq2.8W_0$ at the focus. The handedness of this light is maintained
across the focus, making the field globally chiral. In the weak-field regime, such field will imprint a well-defined initial handedness on the time-dependent chiral wavepacket, which can then be probed by using a loosely-focused, time-delayed circularly polarized $\lambda_p=400$ nm. In the strong-field regime, the field will directly imprint its chirality on the photoelectron wavepacket. 

Another interesting direction relates to the application of synthetic 
chiral light to imprint chirality on achiral molecules, extending the family of proposed methods in literature (see e.g. \cite{Owens:2018aa} and the related discussion in the supplementary \cite{SuppNotes}). Last but not least, 
one can also consider 
the creation of Rydberg composites \cite{Hunter:2021aa} with well-defined handedness, which should allow one to study chiral interactions and the corresponding emergent enantio-sensitive properties from the microscopic to the mesoscopic scale. 
\\
\\
N. M. acknowledges Oleg Kornilov and Andrés Ordó{\~n}ez for helpful discussions. N. M. acknowledges DFG QUTIF grant IV 152/6-2. M. I. acknowledges DFG grant IV 152/10-1. O. S. acknowledges DFG QUTIF grant SM 292/5-2. The authors acknowledge the comments of the referees which greatly helped to improve the manuscript.

\bibliographystyle{unsrt}
\bibliography{biblio}

\end{document}